\RequirePackage{xcolor}
\documentclass[journal,twoside,web]{ieeecolor}
\usepackage{fix-cm}
\usepackage{etex}

\usepackage{dblfloatfix}

\usepackage{nag}

\makeatletter
\@ifpackageloaded{xcolor}{}{%
\usepackage[table,x11names,dvipsnames,svgnames]{xcolor}%
}
\makeatother

\usepackage{colortbl}

\usepackage{graphicx}
\usepackage{wrapfig}

\definecolorset{RGB}{lyft}{}{Red,194,39,36;Sunset,202,53,33;Orange,205,68,20;Amber,200,117,42;Yellow,242,169,52;Citron,186,188,44;Lime,112,159,33;Green,56,139,31;Mint,45,118,56;Teal,52,133,135;Cyan,60,132,202;Blue,55,94,248;Indigo,64,13,247;Purple,115,42,248;Pink,176,25,145;Rose,176,32,75}

\usepackage{cite}

\usepackage{microtype}

\usepackage{array}
\usepackage{multirow}
\usepackage{booktabs}
\usepackage{makecell} 

\ifcsname labelindent\endcsname
\let\labelindent\relax
\fi
\usepackage[inline]{enumitem}

\usepackage{subfig}

\newenvironment{lenumerate}[2][]
{\begin{enumerate}[label=(#2\arabic*),leftmargin=0.2in,itemindent=0.15in,#1]}
{\end{enumerate}}

\setlist*[enumerate,1]{label={\itshape\arabic*)}}

\makeatletter
\newcommand{\paragraphswithstop}{%
\let\copyparagraph\paragraph%
\renewcommand\paragraph[1]{\copyparagraph{##1.}}%
}
\makeatother

\usepackage[framemethod=tikz]{mdframed}

\usepackage{suffix}

\usepackage{environ}

\makeatletter
\newsavebox{\boxifnotempty}
\newcommand{\displayifnotempty}[3]{\sbox\boxifnotempty{#2}\setbox0=\hbox{\usebox{\boxifnotempty}\unskip}%
\ifdim\wd0=0pt
\else
 #1\usebox{\boxifnotempty}#3%
\fi%
}
\newcommand{\ifempty}[2]{\setbox0=\hbox{#1\unskip}%
\ifdim\wd0=0pt%
 #2%
\fi%
}
\newcommand{\ifnotempty}[2]{\setbox0=\hbox{#1\unskip}%
\ifdim\wd0>0pt%
 #2%
\fi%
}
\makeatother

\usepackage{algpseudocode}
\usepackage{algorithm}

\usepackage{scrlfile}

\makeatletter
\newcommand*\newstoreddef[1]{
  \BeforeClosingMainAux{%
    \immediate\write\@auxout{%
      \string\restoredef{#1}{\csname #1\endcsname}%
    }%
  }%
}
\newcommand*{\restoredef}[2]{
  \expandafter\gdef\csname stored@#1\endcsname{#2}%
}
\newcommand*{\storeddef}[1]{
  \@ifundefined{stored@#1}{0}{\csname stored@#1\endcsname}%
}
\makeatother

\NewEnviron{tee}{\BODY\typeout{Marker Tee [start] ^^J \BODY ^^JMaker Tee [end]}}

\input{preamble/math}
\newcommand{\real}[1]{\mathbb{R}^{#1}{}}

\newcommand{\bmat}[1]{\begin{bmatrix}#1\end{bmatrix}}

\newcommand{\transpose}{^\mathrm{T}}

\DeclarePairedDelimiter{\norm}{\lVert}{\rVert}

\providecommand{\cC}{\mathcal{C}}

\providecommand{\cK}{\mathcal{K}}

\providecommand{\cU}{\mathcal{U}}

\usepackage{units}

\newcommand{\newcolorlabel}[2]{%
  \expandafter\newcommand\csname #1\endcsname[1]{%
    \colorbox{#2}{\color{white}\textsf{\textbf{##1}}}}%
}

\newcommand{\newcommenter}[2]{%
  \expandafter\newcommand\csname #1\endcsname[1]{%
    \fcolorbox{#2}{#2}{\color{white}\textsf{\textbf{#1}}}
    {\color{#2}##1}}%
  \expandafter\newcommand\csname at#1\endcsname{%
    \fcolorbox{#2}{#2}{\color{white}\textsf{\textbf{@#1}}}
    {\color{#2}}}%
  \expandafter\newcommand\csname #1hl\endcsname[2]{%
    \colorbox{#2}{\color{white}\textsf{\textbf{#1}}}\sethlcolor{Azure2}\hl{##2}~%
    \expandafter\ifx\csname commentarrow\endcsname\relax$\leftarrow$\else \commentarrow[#2]\fi~%
    {\color{#2}##1}}%
  \expandafter\newcommand\csname #1st\endcsname[2]{%
    \colorbox{#2}{\color{white}\textsf{\textbf{#1}}}\sout{##2}~%
    \expandafter\ifx\csname commentarrow\endcsname\relax$\leftarrow$\else \commentarrow[#2]\fi~%
    {\color{#2}##1}}%
}
\newcommenter{TODO}{DodgerBlue1}
\newcommenter{rtron}{Green3}
\newcommenter{zyang}{blue}

\usepackage{comment}

\usepackage{pdfcomment}

\usepackage{soul}

\usepackage[normalem]{ulem}

\usepackage{csquotes}

\usepackage{tikz}
\usetikzlibrary{calc}
\usetikzlibrary{matrix}
\usetikzlibrary{chains}
\usetikzlibrary{shapes.geometric}
\usetikzlibrary{arrows.meta}
\usetikzlibrary{decorations.pathreplacing}
\usetikzlibrary{backgrounds}

\tikzset{
  dim above/.style={to path={\pgfextra{
        \pgfinterruptpath
        \draw[>=latex,|->|] let
        \p1=($(\tikztostart)!1.5em!90:(\tikztotarget)$),
        \p2=($(\tikztotarget)!1.5em!-90:(\tikztostart)$)
        in(\p1) -- (\p2) node[pos=.5,sloped,above]{#1};
        \endpgfinterruptpath
      }
    }
  },
  dim double above/.style={to path={\pgfextra{
        \pgfinterruptpath
        \draw[>=latex,|->|] let
        \p1=($(\tikztostart)!3em!90:(\tikztotarget)$),
        \p2=($(\tikztotarget)!3em!-90:(\tikztostart)$)
        in(\p1) -- (\p2) node[pos=.5,sloped,above]{#1};
        \endpgfinterruptpath
      }
    }
  },
  dim below/.style={to path={\pgfextra{
        \pgfinterruptpath
        \draw[>=latex,|->|] let 
        \p1=($(\tikztostart)!-1em!-90:(\tikztotarget)$),
        \p2=($(\tikztotarget)!-1em!90:(\tikztostart)$)
        in (\p1) -- (\p2) node[pos=.5,sloped,below]{#1};
        \endpgfinterruptpath
      }
    }
  },
}

\tikzset{
    right angle quadrant/.code={
        \pgfmathsetmacro\quadranta{{1,1,-1,-1}[#1-1]}     
        \pgfmathsetmacro\quadrantb{{1,-1,-1,1}[#1-1]}},
    right angle quadrant=1, 
    right angle length/.code={\def\rightanglelength{#1}},   
    right angle length=2ex, 
    right angle symbol/.style n args={3}{
        insert path={
            let \p0 = ($(#1)!(#3)!(#2)$) in     
                let \p1 = ($(\p0)!\quadranta*\rightanglelength!(#3)$), 
                \p2 = ($(\p0)!\quadrantb*\rightanglelength!(#2)$) in 
                let \p3 = ($(\p1)+(\p2)-(\p0)$) in  
            (\p1) -- (\p3) -- (\p2)
        }
    }
}

\newcommand{\pgfextractangle}[3]{%
    \pgfmathanglebetweenpoints{\pgfpointanchor{#2}{center}}
                              {\pgfpointanchor{#3}{center}}
    \global\let#1\pgfmathresult  
}

\usetikzlibrary{shapes.arrows}
\newcommand{\commentarrow}[1][Azure4]{\tikz[baseline=-3pt]{\node[shape border uses incircle, fill=#1,rotate=180,single arrow, inner sep=1pt, minimum size=6pt, single arrow head extend=2pt]{};}}

\tikzset{ax/.style={-latex,line width=2pt}}

\tikzset{camera/.style={fill=Sienna1,fill opacity=0.5},%
image plane/.style={draw=RoyalBlue3,line width=2pt}}

\usepackage[export]{adjustbox} 
\usepackage[capitalise]{cleveref}

\graphicspath{{image/}}

\DeclareMathAlphabet{\mathcal}{OMS}{cmsy}{m}{n}

\usepackage{lcsys}
\def\BibTeX{{\rm B\kern-.05em{\sc i\kern-.025em b}\kern-.08em
    T\kern-.1667em\lower.7ex\hbox{E}\kern-.125emX}}
\title{Disturbance Observer-Parameterized Control Barrier Function with Adaptive Safety Bounds}

\author{Ziqi Yang$^{1}$, Lihua Xie$^{1}$, \IEEEmembership{Fellow, IEEE}
\thanks{$^{1}$Ziqi Yang and Lihua Xie (Corresponding author) are with School of Electrical and Electronic Engineering, Nanyang Technological University, Singapore 639798
        (email:{\tt\small ziqi.yang@ntu.edu.sg; elhxie@ntu.edu.sg})}
}

\begin{document}

\maketitle
\thispagestyle{empty}
\pagestyle{empty}


\begin{abstract}
  This letter presents a nonlinear disturbance observer-parameterized control barrier function (DOp-CBF) designed for a robust safety control system under external disturbances. This framework emphasizes that the safety bounds are relevant to the disturbances, acknowledging the critical impact of disturbances on system safety. This work incorporates a disturbance observer (DO) as an adaptive mechanism of the safety bounds design. Instead of considering the worst-case scenario, the safety bounds are dynamically adjusted using DO. The forward invariance of the proposed method regardless of the observer error is ensured, and the corresponding optimal control formulation is presented. The performance of the proposed method is demonstrated through simulations of a cruise control problem under varying road grades. The influence of road grade on the safe distance between vehicles is analyzed and managed using a DO.\ The results demonstrate the advantages of this approach in maintaining safety and improving system performance under disturbances.
  \end{abstract}
  

\section{Introduction}
\IEEEPARstart{S}{afety-critical} control is essential in systems where failures can have catastrophic consequences,~\cite{yu2015survey,hsu2023safety,cohen2024safety}. Control Barrier Functions (CBFs)~\cite{Ames2017,Ames2019} have emerged as a tool to enforce safety, and have been applied in various domains~\cite{ames2014control,hsu2015control,glotfelter2017nonsmooth}.
A key challenge for CBF-based controllers is managing disturbances and uncertainties due to reliance on accurate system models. Recent research on robust CBF (RCBF) has focused on introducing robustness into the approaches~\cite{jankovic2018robust,nguyen2021robust,Kolathaya2019,Alan2023,Cohen2022,Xiao2022}. Works like~\cite{Kolathaya2019,Alan2023} propose input-to-state safe CBFs (ISSf-CBFs) that add a relaxation term based on the maximum uncertainty bound. The duality-based method in~\cite{Cohen2022} addresses parametric uncertainty, reducing conservatism through online learning. Further extensions, such as parameter-adaptive CBFs (PACBF) and relaxation-adaptive CBFs (RACBF), improve QP feasibility under time-varying dynamics~\cite{Xiao2022}. A general overview of relaxation-based robust CBF methods is presented in~\cite{Alan2023PCBF}.

\begin{figure}
	\centering
    \subfloat{\includegraphics[width = 0.4\linewidth,valign=c]{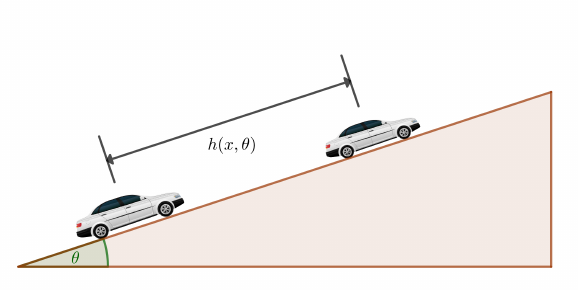}}
    \subfloat{\includegraphics[width=0.6\linewidth,valign=c]{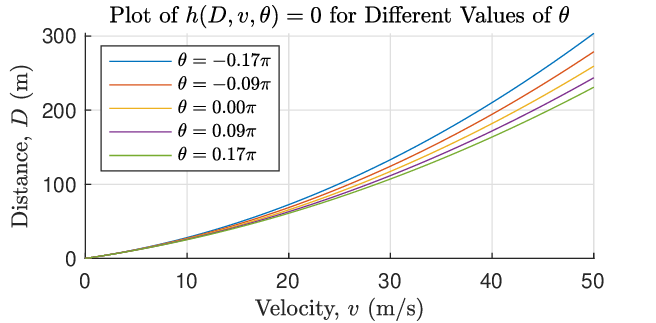}}
    \caption{Due to the impact of road grade on the stopping distance, the safe following distance between vehicles increases on a downhill road and decreases on an uphill road.}\label{fig:example-figure}
\end{figure}
To gain a better understanding of the system states, observer-based techniques have been incorporated into the design of CBF. Some works like~\cite{Devansh2023} utilize the output of the state observer to ensure that the estimated states remain within the safe set, even in the presence of noisy and incomplete measurements. Others adopt disturbance-observer-based control (DOBC)~\cite{Chen2016} to estimate disturbances, a method that has been widely applied across various fields~\cite{jia2023evolver,guo2020multiple,jia2022agile}. Recent works \cite{Alan2023CSL,Sun2024,Das2022,Wang2023,Zinage2023} integrate DOBC with CBFs to enhance robustness. One key challenge of DO-related methods is that the observer error $e_d = d - \hat{d}$ is unobservable, as it depends on the unknown disturbance $d$. Additional concerns arise for DO-CBF methods, as safety must also be guaranteed during the transient phase, before the convergence of the observer errors. Studies like~\cite{Sun2024,Das2022,Wang2023} address this by introducing a tolerance term into the barrier function: $h(x)\to h(x)-\frac{1}{2} e_d^2$. Leveraging the bounded convergence of $e_d$ provided by the disturbance observer design, this modification ensures that the system remains safe even in the presence of relatively large observer errors. 

 While these methods modify the safe set, the underlying safety condition described by $h(x)$ remains unchanged. On the other hand, the critical impact of disturbances on safety design (i.e., $h(x)$) is often overlooked in existing works. For example: wind disturbances affecting a quadrotor (and aircraft) not only challenge its tolerable safe attitude but also dynamically reshape the safe flight corridor based on the winds' magnitude and direction. Similarly, in adaptive cruise control, road condition variations (road type and grade, wet or icy surfaces, etc.) directly affect the breaking distance, requiring dynamic adjustments to the safe following distance (see \cref{fig:example-figure}). In such cases, existing methods either ignore this aspect or rely on worst-case assumptions, leading to overly conservative designs.

In this work, a nonlinear \emph{disturbance observer parameterized control barrier function (DOp-CBF)} framework extends the capabilities of DO-CBF integration by focusing this overlooked aspect. 
Compared with existing DO-CBF~\cite{Sun2024,Das2022,Wang2023}, which introduces a static tolerance term to account for observer error, the proposed method incorporates a dynamic adaptation of the safety set. This is achieved through a two-step modification of the CBF:
\begin{itemize}
  \item \emph{Dynamic disturbance impact}: The safety set is first adjusted to account for the real-time influence of disturbances on the system safety condition.
  \item \emph{Observer error mitigation}: A constant term, inspired by~\cite{Wang2023}, is used to handle observer error, ensuring robust safety during the transient phase.
\end{itemize}

This method is validated on a cruise control problem under various road grades as disturbances. A road grade observer is provided as the adaptation DO term. The results demonstrate the advantages of the proposed method in both bound tightening and relaxing conditions. 

Overall, the main contributions of this letter are as follows:
\begin{itemize}
  \item While most CBF methods focus on a fixed safety set, a novel DOp-CBF is presented in this paper to account for the impact of disturbance on the safety set. This method is the first to consider a disturbance-dependent safety bound, providing online adjustment of barrier function using the estimation from the DO.\ The safety guarantee regardless of the observer error is also provided. 
  \item When the disturbance-dependent term is set to a fixed constant, DOp-CBF can be reduced to a regular DO-CBF.\ Compared with~\cite{Das2022,Wang2023}, DO-CBF in this approach requires only a bound on the time derivative of the disturbance, leading to a less conservative setup.
\end{itemize}

The remainder of this letter is organized as follows. \cref{sec:preliminaries} introduces the background and preliminaries. \cref{sec:problem-formulation} introduces the problem formulation. \cref{sec:DO-pBF} presents the concept of the DOp-CBF. \cref{sec:acc} introduces the application of the proposed methods to the adaptive cruise control (ACC) problem. \cref{sec:simulation} presents the simulation results.\ \cref{sec:conclusion} summarizes this paper.

\section{Preliminaries}\label{sec:preliminaries}
\subsection{Control Barrier Functions}
Consider the dynamics of a nonlinear control affine system of the form
\begin{equation}\label{eq:system_dyn}
  \dot x = f(x) + g(x)u,
\end{equation}
where $x\in \real{n}$ is the system state and $u\in \cU \subset \real{m}$ is the control input, $f:\real{n}\rightarrow \real{n}$, $g \neq 0 : \real{n} \rightarrow \real{n\times m}$ are locally Lipschitz functions. The control input $u = k(x)$ is usually specified as a function of system state $x$, where $k : \real{n}\rightarrow \real{m}$ is locally Lipschitz. System \emph{safety} is usually interpreted as the \emph{forward invariance} of a closed safe set  $C\subset \real{n}$. 
\begin{definition}
  For a set $\cC$ and a closed loop system~\eqref{eq:system_dyn}, any $x(0)\in \cC$, $\cC$ is considered \emph{forward invariant} if $x(t)\in C, \forall t \geq 0$.
\end{definition}

Consider the set $\cC$ defined as a zero-superlevel set of a continuously differentiable function $h : \real{n} \rightarrow \real{}$, 
\begin{equation}\label{eq:barrier-function}
  \cC :=\{x\in \real{n} | h(x)\geq 0 \}.
\end{equation}
\emph{CBF} provides constraints on controller that render the forward invariance (i.e.\ safety) of the safety set $\cC$ as~\eqref{eq:barrier-function}, is defined as follows.
\begin{definition}
  Let set $\cC$ defined by~\eqref{eq:barrier-function}, a continuously differentiable function $h : \real{n} \rightarrow \real{}$ is a CBF for system~\eqref{eq:system_dyn} if there exist an extended class $\cK$ function $\alpha$ such that, for $\forall x \in \cC$, 
  \begin{equation}\label{eq:CBF_condition}
    \sup_{u\in U} L_f h(x) + L_g h(x) u \geq -\alpha(h(x)).
  \end{equation}
\end{definition}


The definition of CBFs captures affine constraints on the control input, allowing control values that satisfy the CBF condition to be computed by solving a quadratic program (QP). Similarly, Control Lyapunov Function (\emph{CLF}) can be formulated into affine constraints on control inputs to guarantee the system stability.
\begin{definition}
  A continuously differentiable positive definite function $V : \real{n} \rightarrow \real{}_{\geq 0}$ is a \emph{CLF} for system~\eqref{eq:system_dyn} if there exists an extended class $\cK$ function $\gamma$ such that, for $\forall x \in \real{n}$, 
  \begin{equation}\label{eq:CLF_condition}
    \inf_{u\in U} L_f V(x) + L_g V(x) u \leq -\gamma(V(x)).
  \end{equation}
\end{definition}

The CBF condition~\eqref{eq:CBF_condition} and CLF condition~\eqref{eq:CLF_condition} are usually formulated as constraints for an optimal control problem and solved via quadratic programming. The online CBF-CLF-QP formulation is formulated as follows:
\begin{align}\label{eq:optimal-control-problem-original}
  \min_{u\in R^m} & \norm{u-k(x)}_2^2\\
  s.t. & L_f V(x) + L_g V(x) u \leq -\gamma(V(x)) + \sigma \\
       & L_f h(x) + L_g h(x) u \geq -\alpha(h(x)),
\end{align}
where $\sigma$ is a relaxation variable to ensure the feasibility of this problem when there is a potential conflict between CBF and CLF constraints.
\section{Problem formulation}\label{sec:problem-formulation}
\begin{figure}
  \centering
  \includegraphics[width = 0.8\linewidth,trim = 0.3cm 0.1cm 0.4cm 0.3cm, clip]{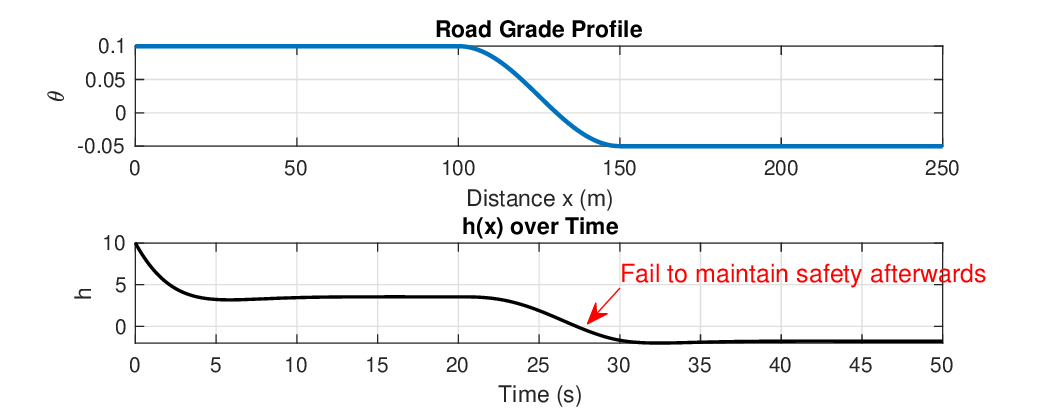}
  \caption{Simulation of an adaptive cruise control problem using regular CBF with road grade considered as disturbance. System safety starts to fail around $t\approx 26s$.}\label{fig:regular-cbf}
\end{figure}

Consider a disturbed nonlinear dynamical system described by the following state-space representation:

\begin{equation}\label{eq:system_dyn_disturbed}
  \dot x = f(x)+ g_1(x)u + g_2(x)d,
\end{equation}
where $d : \real{l}$ is the external disturbance, $g_1 \neq 0$ and $g_2 : \real{n} \rightarrow \real{n\times l}$ are locally Lipschitz functions. Under external disturbance, the regular CBF-CLF controller is unable to maintain safety, as shown in~\cref{fig:regular-cbf}.

To account for the impact of disturbances on the safety condition, the barrier function is designed to also be a function of the disturbance $d$. The disturbance parameterized barrier function $h(x,d)$ incorporates the disturbance into the safety condition, expressed as:
\begin{equation}
  h_d(x,d) = h(x) + \delta (x,d),
\end{equation}
where $\delta: \real{m\times l} \rightarrow \real{}$ is the disturbance parameterized term that modifies the safety boundary in real-time. And the safety set as:
\begin{equation}\label{eq:DO-pBF}
  \cC_{d} = \{x\subset R^n:h_d(x,d)\geq 0\}.
\end{equation}
\begin{problem}
Given the system described as~\eqref{eq:system_dyn_disturbed}, find the control law $u(t)$ such that the system state $x(t)$ remains within the safeset $\cC_d$ for all time $t\geq 0$ despite the presence of unknown disturbances $d$.
\end{problem}
\section{Disturbance observer based safety control design}\label{sec:DO-pBF}
This section presents the concept of the \emph{disturbance observer-parameterized barrier function}. A DO is introduced to estimate the unknown disturbances affecting the system and the output of the DO is utilized as a parameter within the barrier function design. 
\subsection{Disturbance Observer}
DO is a framework to estimate external disturbance $d$ through system variable $x$ when $d$ is not directly available in real-time applications. Assume a slowly varying disturbance with an upper bound on its derivative $\norm{\dot d} \leq \omega$.

The DO considered in this paper is designed as:
\begin{align} \label{eq:DOB}
  \dot z =& -l(x)(f(x)+g_1(x)u+g_2(x)\hat d),\\
  \hat{d} =& z + p(x),
\end{align}
where $\hat d \in \real{l}$ is the estimation of disturbance $d$, $z \in \real{l}$ is the internal state variable of the observer to replace the term $\dot x$, and $p : \real{n}\rightarrow \real{l}$, $l(x)= \frac{\partial p(x)}{\partial x}\neq 0$ are nonlinear functions to be designed for the nonlinear observer. 
The observer error of DO is defined as 
\begin{equation} \label{eq:observer-error}
  e_d = d - \hat d.
\end{equation}
The error dynamics $\dot e_d = \dot d- \dot z - l(x)\dot x$ can be obtained by substituting~\eqref{eq:DOB} into~\eqref{eq:observer-error} 
\begin{align}
  \dot e_d = & \dot d - l(x)(\dot x - f(x)-g_1(x)u-g_2\hat d) \\
   = & \dot d - l(x)(g_2(x) (d - \hat d)) \\
   = & \dot d - l(x) g_2(x) e_d.
\end{align}
Choose a Lyapunov function candidate $V_e = \frac{1}{2}e_d\transpose e_d$. Suppose $l(x)$ is chosen such that $l(x)g_2(x)\leq 1$. Define $\alpha_d = 1-\nu/4$, where $\nu$ is a scalar satisfying~$4\geq \nu \geq 0$. The time derivative of $V_e$ yields
\begin{equation}\label{eq:observer_error_convergence}
  \begin{aligned}
    \dot{V}_e &= e_d\transpose (\dot d - l(x) g_2(x) e_d)  \leq -l(x)g_2(x) e_d\transpose e_d + \norm{e_d}\omega \\  = &-2l(x)g_2(x)V_e - \left\lVert\sqrt{\frac{\nu}{2}}\norm{e_d} - \sqrt{\frac{1}{2\nu}}\omega \right\rVert^2_2+ \frac{\nu}{2} e_d\transpose e_d + \frac{\omega^2}{2\nu} 
    \\  \leq & -2 (l(x)g_2(x)-\frac{\nu}{4}) V_e + \frac{\omega^2}{2\nu} = -2\alpha_d V_e +  \frac{\omega^2}{2\nu}. 
  \end{aligned}
\end{equation}
~\eqref{eq:observer_error_convergence} implies that the estimation $\hat d$ converges to a bounded error exponentially.

By taking~\eqref{eq:observer-error} into~\eqref{eq:system_dyn_disturbed}, the overall closed-system dynamics can be written as:
\begin{equation}\label{eq:system_dyn_do}
  \dot x = f(x)+g_1(x)u+g_2(x)(\hat d + e_d).
\end{equation}


\subsection{DO-parameterized Control Barrier Function}\label{sec:DOCBF}

Existing DO-CBF methods \cite{Sun2024,Das2022,Alan2023CSL,Wang2023} are sufficient for maintaining safety under disturbances. However, when disturbances directly impact safety conditions, these methods can result in overly conservative outcomes. This letter introduces a DO-parameterized barrier function to account for such impacts. By incorporating DO as a \emph{dynamic disturbance impact} term into the safety bounds, this approach expands the control search space, thereby reducing conservatism and allowing for more adaptive safety management. 

The modified barrier function is written as:
\begin{equation}\label{eq:pBF}
  h_{\hat d}(x,\hat d) = h(x) + \delta(x,\hat d)
\end{equation}
where $h(x)$ represents the nominal safety bound in the absence of disturbances, and $\delta : \real{n}\times\real{l} \rightarrow \real{}$ is the dynamic disturbance impact term to account for the safety bound's response under the estimated disturbance $\hat d$. 
The corresponding safety set is designed as 
\begin{equation}\label{eq:DOpbf-safety-set}
  \cC_{\hat d} := \{x\in \real{n} | h_{\hat d}(x) \geq 0 \}.
\end{equation}

Due to the fact that the actual disturbance $d$ is unknown, the observer error $e_d$ is not directly observable. Thus, a perfect compensation is not achievable. To address this issue, an additional \emph{observer error mitigation} term is incorporated to ensure safety.

\begin{definition}\label{definition:max-relaxation}
  Consider the system~\eqref{eq:system_dyn_do}, and DO designed as~\eqref{eq:DOB}, a continuously differentiable function $h_{\hat d}$ is called a disturbance observer-parameterized control barrier function (DOp-CBF) for $\cC_{\hat d}$ defined in~\eqref{eq:DOpbf-safety-set}, if there exists a positive constant $2\alpha_d>\alpha > 0$, such that 
  \begin{multline}\label{eq:dop-cbf-condition}
    \sup_u L_f h_{\hat d}(x) + L_{g_1} h_{\hat d}(x)  u + L_{g_2} h_{\hat d}(x) \hat d \geq\\ -\alpha h_{\hat d}  + \iota_{\hat d}(x),
  \end{multline}
  where $\iota_{\hat d}(x)$ serves as the \emph{observer error mitigation} term to accommodate for the observer error:
  \begin{equation}\label{eq:DOpCBF-robust-term}
    \iota_{\hat d}(x) =\left( \frac{L_{g_2}h_{\hat d}(x) + \frac{\partial \delta}{\partial \hat d} l(x)g_2(x)}{2\sqrt{\sigma\alpha_d- \sigma\alpha /2 } }\right)^2 + \sigma \frac{\omega^2}{2\nu}.
  \end{equation}
\end{definition}
\vspace{0.2cm}
\begin{theorem}\label{theorem:forward-invariant-mCBF}
  For a dynamic system~\eqref{eq:system_dyn_disturbed} with DO designed as~\eqref{eq:DOB}, a DOp-CBF $h_{\hat d}$ defining the safe set $\cC_{\hat d}$~\eqref{eq:DOpbf-safety-set}, a Lipschitz continuous controller $u \in K_{DOpCBF}$ where
  \begin{multline}\label{eq:DOpCBF-control-set}
    K_{DOpCBF}(x,\hat d) = \{u\in \real{m} | L_f h_{\hat d}(x) + L_{g_1} h_{\hat d}(x) u \\
    + L_{g_2} h_{\hat d}(x)\hat d \geq -\alpha h_{\hat d}(x) +  \iota_{\hat d}(x) \},
  \end{multline}
   will render $\cC_{\hat d}$ forward invariant for all $d$ satisfying $\norm{\dot d} < \omega$.
\end{theorem}

\begin{proof}
  To accommodate for the observer error, inspired by \cite{Wang2023,Sun2024} a robustness term is added to the CBF candidate $h_{\hat d e} = h_{\hat d} - \sigma V_e$.  The corresponding safety set is designed as 
\begin{align}\label{eq:dopbf-e}
  \cC_{\hat d e} :=& \{x\in \real{n} | h_{\hat d e}(x) \geq 0 \},\\
  \partial \cC_{\hat d e}:=& \{x\in \real{n} | h_{\hat d e}(x)  = 0 \},\label{eq:dopbf_edge}\\
  Int(\cC_{\hat d e}):=& \{x\in \real{n} | h_{\hat d e}(x) > 0 \}.
\end{align}
The derivative of $h_{\hat d e}$ can be derived as:
\begin{equation}\label{eq:dot_h_e}
  \begin{aligned}
    \dot h_{\hat d e}(x) = & \frac{\partial h_{\hat d}(x)}{\partial x}\dot x + \frac{\partial \delta}{\partial \hat d} \dot{\hat{d}}- \sigma \dot V_e \\
     =& L_f h_{\hat d}(x)+ L_{g_1}h_{\hat d}(x) u + L_{g_2}h_{\hat d}(x)\hat d  + L_{g_2}h_{\hat d}(x)e_d \\
     & +\frac{\partial \delta}{\partial \hat d} l(x)g_2(x) e_d - \sigma \dot V_e.\\
  \end{aligned}
\end{equation}
\eqref{eq:observer_error_convergence} provide an upper bound for $\dot{V}_e \leq - 2\alpha_d V_e + \frac{\omega^2}{2\nu}$. A lower bound for $\dot h_{\hat d e}(x)$ can be derived as:
\begin{equation}\label{eq:dot_h_lower_bound}
  \begin{aligned}
    \dot h_{\hat d e}(x) \geq & L_f h_{\hat d}(x) + L_{g_1} h_{\hat d}(x) u + L_{g_2} h_{\hat d}(x)\hat d +\Big(L_{g_2}h_{\hat d}(x)\\
     & +\frac{\partial \delta}{\partial \hat d} l(x)g_2(x) \Big)e_d + \sigma \left(2\alpha_d V_e - \frac{\omega^2}{2\nu}\right).
  \end{aligned}
\end{equation}

First consider the boundary of the safe set $x\in\cC_{\hat d e}$~\eqref{eq:dopbf_edge}, $h_{\hat d e}(x) =0$, thus $h_{\hat d}=\sigma V_e$.
Controller in~\eqref{eq:dop-cbf-condition} yields
  \begin{equation}\label{eq:docbf-condition-boundary}
    L_f h_{\hat d}(x) + L_{g_1} h_{\hat d}(x) u + L_{g_2} h_{\hat d}(x)\hat d \geq -\alpha \sigma V_e  + \iota_{\hat d}(x).
  \end{equation}
Thus, corresponding controller $u\in K_{DOpCBF}(x,\hat d)$ renders
\begin{equation}
    \begin{aligned}\label{eq:docbf-proof}
          \dot h_{\hat d e} & \geq -\alpha\sigma V_e + \iota_{\hat d}(x) + \Big(L_{g_2}h_{\hat d}(x) +\frac{\partial \delta}{\partial \hat d} l(x)g_2(x) \Big)e_d \\
          &  + \sigma \left(\alpha_d e^T e  - \frac{\omega^2}{2\nu}\right)\\
           &=(\sigma\alpha_d - \frac{\sigma\alpha}{2}) e^T e + \Big(L_{g_2}h_{\hat d}(x) +\frac{\partial \delta}{\partial \hat d} l(x)g_2(x) \Big)e_d \\
           &-\sigma \frac{\omega^2}{2\nu}+ \iota_{\hat d}(x).
    \end{aligned}
\end{equation}
To ensure the non-negativity of $\dot h_{\hat d e}$, taking the value of $\iota_{\hat d}(x)$ from~\eqref{eq:DOpCBF-robust-term} into~\eqref{eq:docbf-proof} yields:
  \begin{equation}
    \begin{aligned}
      \dot h_{\hat d e} & \geq  (\sigma\alpha_d - \frac{\sigma\alpha}{2}) e^T e + \Big(L_{g_2}h_{\hat d}(x) +\frac{\partial \delta}{\partial \hat d} l(x)g_2(x) \Big)e_d \\
        & -\sigma \frac{\omega^2}{2\nu}+ \left( \frac{L_{g_2}h_{\hat d}(x) + \frac{\partial \delta}{\partial \hat d} l(x)g_2(x)}{2\sqrt{\sigma\alpha_d- \sigma\alpha /2 } }\right)^2 + \sigma \frac{\omega^2}{2\nu}\\
        & =\left\lVert \sqrt{\sigma\alpha_d-\frac{\sigma \alpha}{2}} e + \frac{L_{g_2}h_{\hat d}(x)+\frac{\partial \delta}{\partial \hat d} l(x)g_2(x)}{2\sqrt{\sigma\alpha_d- \sigma\alpha /2 } } \right\rVert_2^2\\
        & \geq 0.
    \end{aligned}
\end{equation}
It is indicated that for any $x\in\partial\cC_{\hat d e}$,  $u\in K_{DOpCBF}(x,\hat d)$ will render $\dot h_{\hat d e}(x)\geq 0$ regardless of the observer error $e_d$.\ Thus, according to Nagumo's definition~\cite{blanchini2008set}, set $\cC_{\hat d e}$ is forward invariant, thus set $\cC_{\hat d}$ is forward invariant. 
\end{proof}

\begin{remark}\label{rmk:sigma_choice}
 According to \eqref{eq:observer_error_convergence}, $\hat d$ will converge to a bounded error exponentially, making the subset $\cC_{\hat d e}$ arbitrarily close to the original safe set $\cC_{\hat d}$. A sufficiently small $\sigma$ reduces the distance between these two sets but also causes the first term in the safety condition~\eqref{eq:DOpCBF-robust-term} to become very large, thereby increasing the conservativeness of the allowable control set~\eqref{eq:DOpCBF-control-set}. The trade-off in selecting  $\sigma$ must be carefully balanced to ensure both safety and the practical feasibility of the control law.
\end{remark}
\begin{remark}
\cref{theorem:forward-invariant-mCBF} provides a general framework for attenuating the influence of disturbances on the safety properties of the system. By introducing additional compensation term $-\frac{g_2(x)}{g_1(x)} \hat d$ into the controller,~\eqref{eq:dop-cbf-condition} can be specialized to represent the safety properties of DOBC related techniques.
\end{remark}

\begin{remark}\label{rmk:DO-CBF}
  When $\delta(x,\hat d) := 0$, the forward invariance condition of DOp-CBF in \cref{theorem:forward-invariant-mCBF} is reduced to a similar the DO-CBF conditions introduced in~\cite{Wang2023,Sun2024}.
\end{remark}

\section{Application of DOp-CBF for Adaptive Cruise Control}\label{sec:acc}
This section aims to illustrate the practical application of DOp-CBF method by applying it to the ACC problems\cite{ames2014control,Xiao2022}. The goal is to demonstrate how the proposed approach effectively handles the challenges of maintaining safety in the presence of dynamic disturbances.

\subsubsection*{Vehicle Dynamics} Consider a state $x = [D,v]\transpose$, where $D$ denotes the distance between the controlled vehicle and a preceding vehicle along the same lane and $v$ denotes the speed of the controlled vehicle. Let $u$ be the acceleration of the following vehicle which acts as the control input. Consider a case where there exists a preceding vehicle with a constant speed $v_l$, the following vehicle dynamics can be written as:
  \begin{equation}\label{eq:vehicle_dyn}
    \underbrace{\bmat{\dot D\\\dot v}}_{\dot x} = \underbrace{\bmat{v_l - v\\ - \frac{cv^2}{M}}}_{f(x)} + \underbrace{\bmat{0 \\ \frac{1}{M}}}_{g_1(x)} u + \underbrace{\bmat{0\\\frac{1}{M}}}_{g_2(x)} a(\theta)
  \end{equation}
  where $c$ denotes air drag coefficient, $M$ denotes the mass of the vehicle and $\theta$ denotes the time-varying road grade considered as the external disturbance, and $a(\theta) = g\sin\theta$ with $g$ the gravitational acceleration. 
  \subsubsection*{Road grade observer}
  With respect to the system dynamics, consider a slowly varying road grade $\dot \theta = 0$, a \emph{road-grade observer} is constructed to estimate the portion of gravitational acceleration that contributes to the vehicle's dynamics along the road:
  \begin{equation}\label{eq:road-grade-observer}
    \left\{
    \begin{aligned}
      \dot \xi =& -L_r(f(x)+g_1(x)u+g_2(x)\hat d),\\
      \hat \theta =& \arcsin( \sin(\xi + L_r x)),\\
      \hat{d} =& a(\theta),
    \end{aligned}
    \right.
  \end{equation}
  where $L_r = \bmat{l_1, l_2}$ is the observer gain, and term $\arcsin(\sin(r))$ is used to keep the observed angle $r$ within $\bmat{0,\pi/2}$.
  \begin{remark}
    As demonstrated in some DO-related experiments~\cite{jia2022agile}, when disturbance frequency is significantly lower than control frequency, low-frequency components dominate with negligible derivatives due to their smooth continuity. Therefore, it is reasonable to treat $\dot{\theta} = 0$ in the simulation scenario.
  \end{remark}

  \subsubsection*{Vehicle objectives}
  The vehicle is required to minimize acceleration while attempting to maintain a prescribed reference speed at $v_r$. Following the dynamic of the second state $v$ in~\eqref{eq:vehicle_dyn}, the minimization of acceleration is written as the objective function in the form of 
  \begin{equation}
    \min_u \left(\frac{u - cv^2}{M}\right).
  \end{equation}
  
  The reference velocity $v_r$ is constructed as a CLF with $V_r(x) = {(v-v_r)}^2$ with corresponding CLF constraints written as:
  \begin{equation}
    L_f V_r(x) + L_{g_1}V_r(x) u + L_{g_2}V_r(x) \hat d \leq -\gamma V_r.
  \end{equation}

  \subsubsection*{Safety constraints with DOp-CBF}
  To ensure safety, the distance $D$ between the controlled vehicle and the closest preceding vehicle must exceed a minimum safe distance which is usually evaluated by the sum of braking distance and reaction distance. Therefore, the barrier function is defined as:
  \begin{equation}
    h_{\hat \theta}(x,\hat \theta) = D-D_{sf}(\hat \theta) - Tv,
  \end{equation}
  where $T>0$ is the reaction time, and
  \begin{equation}
      D_{sf}(\hat \theta) = \frac{v^2}{2(\mu+\sin{\hat \theta})g},
  \end{equation}
 represents the estimated braking distance of the controlled vehicle, accounting for the road grade disturbance as observed by~\eqref{eq:road-grade-observer}, $\mu$ constant representing the road-tire adhesion coefficient (assuming road conditions are constant). 
 
  The corresponding DOp-CBF constraint yields:
\begin{equation}\label{eq:acc-dop-cbf}
  L_f h_{\hat \theta}(x) + L_{g_1} h_{\hat \theta}(x) u \\
    + L_{g_2} h_{\hat \theta}(x)\hat d \geq -\alpha h_{\hat \theta}(x) + \iota_{\hat d}(x),
\end{equation}
with 
\begin{equation}
  \iota_{\hat d}(x) = \left( \frac{L_{g_2}h_{\hat d}(x) + \frac{\partial \delta}{\partial \hat d} l(x)g_2(x)}{2\sqrt{\sigma\alpha_d- \sigma\alpha /2 } }\right)^2.
\end{equation}

\section{Implementation and results}\label{sec:simulation}
In this section, the simulation of the ACC problem introduced in \cref{sec:acc} under varying road conditions is presented. The DOp-CBF in \cref{sec:acc} is implemented in MATLAB to illustrate the method proposed in \cref{sec:DO-pBF}. The control problem is formulated as QP and solved via \emph{quadprog} and the dynamics is solved via \emph{ode45}.

\begin{figure*}
  \centering
  \subfloat{\includegraphics[width = 0.7\linewidth,valign=c]{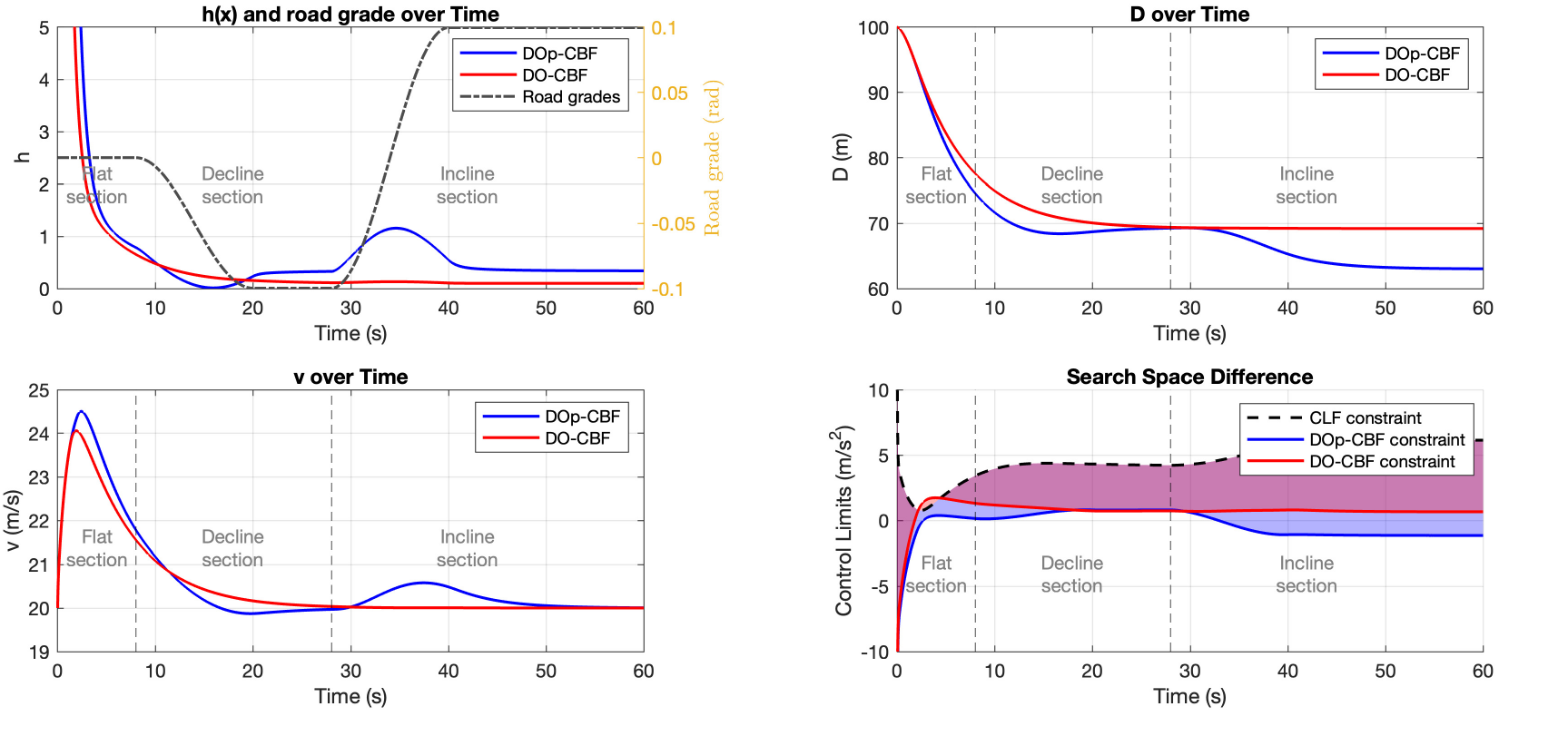}}
  \subfloat{\includegraphics[width = 0.3\linewidth,valign=c]{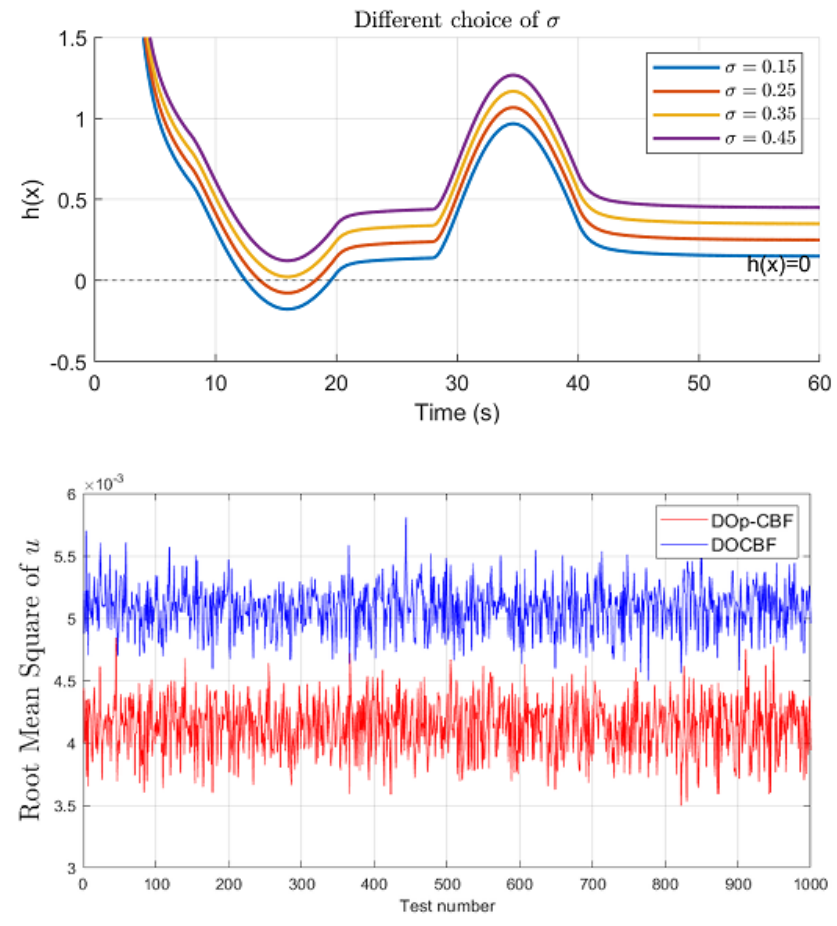}}
  \caption{(Left) Simulation result of a road with three sections, figure indicating $h(x)$ value and road grade with respect to time, following distance, following speed and search space difference. (Top right) Results for different choice of $\sigma$. (Bottom right) Simulation result for RMS of $u$ under 1000 random road tests. }\label{fig:result-compare-2}
\end{figure*}

The parameters are set as follows: $g = 9.81 m/s^2$, $\gamma = 0.006$, $c = 9,99428 1/m$, $T = 2 s$, $M = 1650 kg$, $\mu = 0.8$. The initial conditions are set as $v(0) = 20m/s$, $D(0) = 70m$. Consider the desired speed is at $v_r = 25m/s$, and the leading vehicle is at constant speed $v_l = 20m/s$, and the fixed safe following distance is set to $D_{sf}=25m$. The road grade observer gain is set as $L = \bmat{3,3}\transpose$. The disturbance road grade is shown as yellow dashed line in \cref{fig:result-compare-2}.

Simulation results in~\cref{fig:regular-cbf} show that the regular CBF controller, without considering road grade disturbance, cannot guarantee safety. The road grade observer based DOp-CBF~\eqref{eq:acc-dop-cbf} is implemented to compensate for the varying road grade and DO-CBF in \cite{Wang2023} is implemented as a comparison. Since disturbance is not considered in the design of the barrier function, the safe following distance needs to consider the worst-case scenario (i.e.\ the steepest downhill). Thus, the DO-CBF barrier function is designed as:
 \begin{equation}
  h_{sd}(x) = D- \frac{v^2}{2g(\mu+\sin\theta_{dm})}- Tv, 
 \end{equation}
where $\theta_{dm}$ indicates the maximum decline road grade. 

The simulation results are shown in~\cref{fig:result-compare-2}. Both DO-CBF and DOp-CBF are tested on a road containing three sections, a flat section, a decline section and an incline section. As shown in~\cref{fig:result-compare-2}, both methods can maintain the safety (i.e.\ keeping $h(x)\geq 0$) during the entire simulation period. Compared with DO-CBF, the DOp-CBF provides a more flexible design of the relative safety distance and is able to adjust the barrier function adaptively. The decline section requires a larger safe following distance. Without enough information, DO-CBF needs to be conservative by keeping the following distance under the maximum decline road condition around $70m$ (red line in~\cref{fig:result-compare-2}). Meanwhile, by utilizing the estimation of DO, DOp-CBF (blue line in~\cref{fig:result-compare-2}) is able to keep a closer distance during the flat section and gradually adjust the following distance during the incline and decline section.

 As mentioned in~\cref{rmk:sigma_choice}, the choice of $\sigma$ affects both the system performance and safety. As shown in top right of~\cref{fig:result-compare-2}, a small $\sigma$ may underestimate the effect of observer error, increasing the risk of safety failure before the disturbance observer converges. A sufficiently large observer gain $l(x)$ can accelerate DO convergence and reduce the steady-state error. However, it also amplifies noise in the system. On the other hand, a large $\sigma$ may overestimate the impact of the disturbance, leading to a more conservative control result. Therefore, $\sigma$ and $l(x)$ should be carefully selected to balance noise sensitivity and robustness. 

Due to the requirement of a larger $\theta$, DOp-CBF needs to operate at a slightly greater distance from the edge of the safe set compared to DO-CBF (\cref{fig:result-compare-2} left). However, DOp-CBF renders a larger safe set most of the time, enabling a larger search space during optimal control computation (see \cref{fig:result-compare-2} middle, red fillings indicate overlapping search space and blue fillings indicate additional search space provided by DOp-CBF). This flexibility offers several benefits, including the potential for improved performance, avoidance of conflicts with the CLF (as observed in the DO-CBF simulation), and increased adaptability.

 To compare the effectiveness of DO-CBF and DOp-CBF, the Root Mean Square (RMS) of the derivative of the control input $u$ quantifies its smoothness, serving as an evaluation metric for comfort, is defined as $\text{RMS} = \sqrt{\frac{1}{T} \int_0^T \left( \frac{du(t)}{dt} \right)^2 dt}.$
After the initial convergence of DO, the RMS for DOp-CBF controller output is improved by $11.96\%$ (compared with DO-CBF, from 0.0038 to 0.0034), thanks to the extra search space provided by the adaptive safety bound.  To further validate the approach, 1,000 randomized road tests were conducted with random varying road grades between $-0.2 Rad$ and $0.2 Rad$ with an upper bound on $\dot{\theta}$. In all cases, safety was successfully maintained. The RMS compensation results indicate that on average, smoothness improved by $22.73\%$, with a maximum improvement of $42.18\%$ and minimal improvement of $3.60\%$.



\section{Conclusion}\label{sec:conclusion}
This paper presented DOp-CBF to enhance the robustness of safety-critical control systems under external disturbances. The approach allows the safety bounds to be dynamically adjusted based on real-time disturbance estimates, provided by DO. For disturbed systems, DOp-CBF has shown more flexibility and adaptability in safety bound design compared to traditional methods, as it is not required to consider the worst-case scenario. The effectiveness of the DOp-CBF method has been demonstrated through a simulation of an ACC problem, focusing on the influence of road grade on safe distance maintenance. 

However, limitations still existed, mainly focusing on the reliance of disturbance observer design on model accuracy and its ability to handle rapidly changing disturbances. These issues will be addressed in future work. Future work will also focus on exploring its applicability in real-world scenarios with a particular emphasis on quadrotors under wind disturbances. Furthermore, the dynamic adjustment of the relaxation variable $\sigma$ is also a promising area for future research.

{
\bibliographystyle{IEEEtran}
\bibliography{Biblio/thesis,Biblio/ADMM_planning,Biblio/tron,Biblio/CBF,Biblio/CBF_uncertainty,Biblio/DOB}

\begin{thebibliography}{10}
\providecommand{\url}[1]{#1}
\csname url@samestyle\endcsname
\providecommand{\newblock}{\relax}
\providecommand{\bibinfo}[2]{#2}
\providecommand{\BIBentrySTDinterwordspacing}{\spaceskip=0pt\relax}
\providecommand{\BIBentryALTinterwordstretchfactor}{4}
\providecommand{\BIBentryALTinterwordspacing}{\spaceskip=\fontdimen2\font plus
\BIBentryALTinterwordstretchfactor\fontdimen3\font minus
  \fontdimen4\font\relax}
\providecommand{\BIBforeignlanguage}[2]{{%
\expandafter\ifx\csname l@#1\endcsname\relax
\typeout{** WARNING: IEEEtran.bst: No hyphenation pattern has been}%
\typeout{** loaded for the language `#1'. Using the pattern for}%
\typeout{** the default language instead.}%
\else
\language=\csname l@#1\endcsname
\fi
#2}}
\providecommand{\BIBdecl}{\relax}
\BIBdecl

\bibitem{yu2015survey}
X.~Yu and J.~Jiang, ``A survey of fault-tolerant controllers based on
  safety-related issues,'' \emph{Annual Reviews in Control}, vol.~39, pp.
  46--57, 2015.

\bibitem{hsu2023safety}
K.-C. Hsu, H.~Hu, and J.~F. Fisac, ``The safety filter: A unified view of
  safety-critical control in autonomous systems,'' \emph{Annual Review of
  Control, Robotics, and Autonomous Systems}, vol.~7, pp. 47--72, 2023.

\bibitem{cohen2024safety}
M.~H. Cohen, T.~G. Molnar, and A.~D. Ames, ``Safety-critical control for
  autonomous systems: Control barrier functions via reduced-order models,''
  \emph{Annual Reviews in Control}, vol.~57, p. 100947, 2024.

\bibitem{Ames2017}
A.~D. Ames, X.~Xu, J.~W. Grizzle, and P.~Tabuada, ``Control barrier function
  based quadratic programs for safety critical systems,'' \emph{IEEE
  Transactions on Automatic Control}, vol.~62, no.~8, pp. 3861--3876, 2017.

\bibitem{Ames2019}
A.~D. Ames, S.~Coogan, M.~Egerstedt, G.~Notomista, K.~Sreenath, and P.~Tabuada,
  ``Control barrier functions: Theory and applications,'' in \emph{2019 18th
  European Control Conference (ECC)}, 2019, pp. 3420--3431.

\bibitem{ames2014control}
A.~D. Ames, J.~W. Grizzle, and P.~Tabuada, ``Control barrier function based
  quadratic programs with application to adaptive cruise control,'' in
  \emph{53rd IEEE Conference on Decision and Control}, 2014, pp. 6271--6278.

\bibitem{hsu2015control}
S.-C. Hsu, X.~Xu, and A.~D. Ames, ``Control barrier function based quadratic
  programs with application to bipedal robotic walking,'' in \emph{2015
  American Control Conference (ACC)}, 2015, pp. 4542--4548.

\bibitem{glotfelter2017nonsmooth}
P.~Glotfelter, J.~Cort{\'e}s, and M.~Egerstedt, ``Nonsmooth barrier functions
  with applications to multi-robot systems,'' \emph{IEEE Control Systems
  Letters}, vol.~1, no.~2, pp. 310--315, 2017.

\bibitem{jankovic2018robust}
M.~Jankovic, ``Robust control barrier functions for constrained stabilization
  of nonlinear systems,'' \emph{Automatica}, vol.~96, pp. 359--367, 2018.

\bibitem{nguyen2021robust}
Q.~Nguyen and K.~Sreenath, ``Robust safety-critical control for dynamic
  robotics,'' \emph{IEEE Transactions on Automatic Control}, vol.~67, no.~3,
  pp. 1073--1088, 2021.

\bibitem{Kolathaya2019}
S.~Kolathaya and A.~D. Ames, ``Input-to-state safety with control barrier
  functions,'' \emph{IEEE Control Systems Letters}, vol.~3, no.~1, pp.
  108--113, 2019.

\bibitem{Alan2023}
A.~Alan, A.~J. Taylor, C.~R. He, A.~D. Ames, and G.~Orosz, ``Control barrier
  functions and input-to-state safety with application to automated vehicles,''
  \emph{IEEE Transactions on Control Systems Technology}, vol.~31, no.~6, pp.
  2744--2759, 2023.

\bibitem{Cohen2022}
M.~H. Cohen, C.~Belta, and R.~Tron, ``Robust control barrier functions for
  nonlinear control systems with uncertainty: A duality-based approach,'' in
  \emph{2022 IEEE 61st Conference on Decision and Control (CDC)}, 2022, pp.
  174--179.

\bibitem{Xiao2022}
W.~Xiao and C.~Belta, ``High-order control barrier functions,'' \emph{IEEE
  Transactions on Automatic Control}, vol.~67, no.~7, pp. 3655--3662, 2022.

\bibitem{Alan2023PCBF}
A.~Alan, T.~G. Molnar, A.~D. Ames, and G.~Orosz, ``Parameterized barrier
  functions to guarantee safety under uncertainty,'' \emph{IEEE Control Systems
  Letters}, vol.~7, pp. 2077--2082, 2023.

\bibitem{Devansh2023}
D.~R. Agrawal and D.~Panagou, ``Safe and robust observer-controller synthesis
  using control barrier functions,'' \emph{IEEE Control Systems Letters},
  vol.~7, pp. 127--132, 2023.

\bibitem{Chen2016}
W.~H. Chen, J.~Yang, L.~Guo, and S.~Li, ``Disturbance-observer-based control
  and related methods - an overview,'' \emph{IEEE Transactions on Industrial
  Electronics}, vol.~63, no.~2, pp. 1083--1095, 2016.

\bibitem{jia2023evolver}
J.~Jia, W.~Zhang, K.~Guo, J.~Wang, X.~Yu, Y.~Shi, and L.~Guo, ``Evolver: Online
  learning and prediction of disturbances for robot control,'' \emph{IEEE
  Transactions on Robotics}, vol.~40, pp. 382--402, 2024.

\bibitem{guo2020multiple}
K.~Guo, J.~Jia, X.~Yu, L.~Guo, and L.~Xie, ``Multiple observers based
  anti-disturbance control for a quadrotor uav against payload and wind
  disturbances,'' \emph{Control Engineering Practice}, vol. 102, p. 104560,
  2020.

\bibitem{jia2022agile}
J.~Jia, K.~Guo, X.~Yu, L.~Guo, and L.~Xie, ``Agile flight control under
  multiple disturbances for quadrotor: Algorithms and evaluation,'' \emph{IEEE
  Transactions on Aerospace and Electronic Systems}, vol.~58, no.~4, pp.
  3049--3062, 2022.

\bibitem{Alan2023CSL}
A.~Alan, T.~G. Molnar, E.~Das, A.~D. Ames, and G.~Orosz, ``Disturbance
  observers for robust safety-critical control with control barrier
  functions,'' \emph{IEEE Control Systems Letters}, vol.~7, pp. 1123--1128,
  2023.

\bibitem{Sun2024}
J.~Sun, J.~Yang, and Z.~Zeng, ``Safety-critical control with control barrier
  function based on disturbance observer,'' \emph{IEEE Transactions on
  Automatic Control}, vol.~69, no.~7, pp. 4750--4756, 2024.

\bibitem{Das2022}
E.~Daş and R.~M. Murray, ``Robust safe control synthesis with disturbance
  observer-based control barrier functions,'' in \emph{2022 IEEE 61st
  Conference on Decision and Control (CDC)}, 2022, pp. 5566--5573.

\bibitem{Wang2023}
Y.~Wang and X.~Xu, ``Disturbance observer-based robust control barrier
  functions,'' in \emph{2023 American Control Conference (ACC)}, 2023, pp.
  3681--3687.

\bibitem{Zinage2023}
V.~Zinage, R.~Chandra, and E.~Bakolas, ``Disturbance observer-based robust
  integral control barrier functions for nonlinear systems with high relative
  degree,'' in \emph{2024 American Control Conference (ACC)}, 2024, pp.
  2470--2475.

\bibitem{blanchini2008set}
F.~Blanchini, S.~Miani \emph{et~al.}, \emph{Set-theoretic methods in
  control}.\hskip 1em plus 0.5em minus 0.4em\relax Springer, 2008, vol.~78.

\end{thebibliography}
}
\end{document}